IAC-22-A1.5.2

# Radiation protection and shielding materials for crewed missions on the surface of Mars

**Dionysios Gakis[a]\*, Dimitra Atri[b]**

[a] *Department of Physics, University of Patras, Patras, Rio, 26504, Greece*, dgakis@upnet.gr
[b] *Center for Space Science, New York University Abu Dhabi, PO Box 129188, Abu Dhabi, UAE*, atri@nyu.edu
\* Corresponding Author

**Abstract**

A potential crewed mission to Mars would require us to solve a number of problems, including how to protect astronauts against the devastating effects of energetic charged particles from Solar and Galactic sources. The radiation environment on Mars is of particular interest, since maintaining optimal absorbed doses by astronauts is crucial to their survival. Here, we give an overview of the conditions on Mars, as determined by theoretical models and in-situ measurements, and present the main proposed strategies to mitigate radiation exposure while on Mars. Specifically, we focus on the passive shielding technique. Several widely used materials, along with some innovative ones and combinations of those, are studied for their behavior against Solar Energetic Particle Events and Galactic Cosmic Rays in the Martian environment. For that purpose, we implement the GEANT4 package, a Monte-Carlo numerical model developed by CERN, which is specifically applied to simulate interactions of radiation with matter. A description of our model will be given, followed by outputs of the numerical model. We conclude that hydrogen-rich materials act as better attenuators, as expected, but other materials can be helpful against cosmic rays too.

**Acronyms/Abbreviations**
SEPs: Solar Energetic Particles
GCR: Galactic Cosmic Rays
LET: Linear Energy Transfer
HZE: High Energy and Charge

## 1. Introduction

The first crewed missions to the red planet are planned for the coming decades. An important aspect of this mission is radiation, as it is considered one of the major challenges for astronaut health [1]. For that purpose, the understanding of the radiation environment on Mars and the adoption of protection measures is needed to ensure the safety of the crew.

Prior to crewed missions to Mars, precursor missions have been sent to quantify the characteristics of the Martian radiation environment. Specifically, the MSL RAD (Mars Science Laboratory Radiation Assessment Detector) instrument measures since 2012 the charged particle and neutral particle radiation, giving the total dose rate and particle spectra from ~10 to hundreds MeV/u [2]. In addition to that, there is an ongoing global effort to use coding models to characterize the radiation environment and validate the measurements by RAD, in order to test the feasibility of a human exploration of Mars.

Astrophysical energetic particles are categorized according to their origin into SEPs and GCR. SEPs are mainly protons associated with impulsive solar flares or gradual coronal mass ejections [3]. On the other hand, GCR enter the Solar System after being accelerated by sources like supernova remnants [4]. At periods of maximum solar activity, GCR fluence are modulated by the interplanetary magnetic field and the solar wind [5]. Likewise, a peak of the GCR intensity is observed when the number of sunspots in the solar photosphere reaches its minimum in the 11-year solar cycle.

Both types of radiation hit the planet Mars. Unlike Earth, Mars' natural shielding is minimal. Its thin $CO_2$-dominated atmosphere and localized mini-magnetosphere cannot attenuate or deflect charged and energetic particles at a great extent (for the most part only the low-energy particles are stopped). As a result, the primary solar and galactic particles interact with the molecules of the atmosphere, producing a cascade of secondaries, which in turn have enough energies to create further particles reaching the surface. Clearly, the radiation field on the surface of Mars is a combination of incoming particles and albedo ones, backscattered by the Martian soil (e.g. [6]).

The radiation environment is not that dangerous on the surface of Mars on its own. For short stays, radiation exposure by SEPs and GCR is calculated to be lower than the currently acceptable dose limits (e.g. [7]). This is attributed to the shielding of Mars, albeit weak. Nevertheless, a typical journey to Mars takes some months, during which astronauts would not be adequately protected (as validated e.g. by the transit of RAD to Mars, [8]). Hence, these dangerous transits, which contribute to the total radiation dose absorbed by their bodies make it necessary to try to mitigate radiation exposure while on Mars too. When radiation accumulates in large parts, it has the potential to cause serious health issues.

The radiation shielding strategies could be divided in operational and engineering. Operational countermeasures include the astronauts' protection during their mission by benefitting from each environment's specific properties. For example, exposure to astronauts may depend on which region of Mars a habitat occupies. Geological structures like cliffs or craters block some part of the incoming radiation [9]. Likewise, the secondary neutron abundancies can vary through different soil composition or variations of protective ice, which is found in some regions near the poles of Mars [10]. We also note the personal protection strategies astronauts follow in order to reduce doses. This can be done by limiting the time outside of the habitat or not wearing their spacesuits as little possible for instance.

The concept of engineering countermeasures lies in structures or tools aiming to protect astronauts from radiation. To this day, the easiest form of such protection is the passive shielding technique, which suggests putting enough radiation absorbing mass between the astronaut and space. However, one of the biggest challenges every space mission has to overcome is the mass limitation. The more massive a spacecraft is, the more difficult and expensive it is to be launched, or the fewer critical items it can take with it. Consequently, it is critical to wisely choose materials that are light to carry, and at the same time stop radiation within their mass efficiently.

The aim of the current study is to classify certain materials in accordance with their radiation shielding effectiveness in the Martian radiation field. This was achieved by creating a numerical model of the conditions of Mars. In that model, we defined a layer of absorbing material protecting a human phantom. By calculating the radiation levels within the phantom, we managed to assess the effectiveness of each material.

In the next section we outline our computational procedure. We then describe our results and discuss their significance. We conclude with a brief summary.

**2. Method**

We performed numerical simulations of the Martian radiation environment using the computational model Geant4 [11]. Geant4 is a toolkit for the modelling of the propagation of radiation in matter. Among other uses, Geant4 is widely used in studies about the interaction of charged cosmic particles with the Martian environment and soil (e.g. [12,13,14,15]) and has been shown to provide consistent results with the MSL RAD measurements [16].

The model we built to represent Mars has been described in [17]. Essentially, a source beam passes through successive layers of atmosphere, the in-question shielding material, the astronaut's body (this is represented by a water slab) and a final layer of Martian regolith. A visualization of our model is provided in Fig. 1. We compute the LET (energy deposited per depth) and the mean energy deposited in the water layer. A material which lets the least energy to pass through the human phantom is considered as radiation effective material.

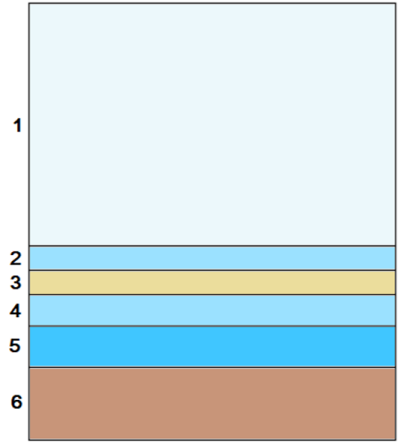

Fig. 1. Martian coding model. It consists of 6 successive layers: Layer 1 (Condensed atmosphere), 2 (Surface atmosphere), 3 (Shielding material), 4 (Surface atmosphere), 5 (Water) and 6 (Soil). Primary radiation hits at the top of the layers. Figure reproduced from [17].

We considered both Sun-originated particles (SEPs) and particles originated outside of the Solar System (GCR). The model by a typical solar event, the one during September 1989 [18], was used to generate the primary particles in the first case, whereas we adopted the Badhwar–O'Neill 2010 model [19] for the latter.

As this is a quite simple model, a slab-plane layered model, we had to validate its accuracy before getting any results and making conclusions. Therefore, we compared our outcomes with the existing ones, deduced by much more complex models and the RAD measurements. In our model, the dose rate by GCR ions in the water layer was found to be about 173 μGy/day during solar maximum conditions and up to almost 300 μGy/day during solar minimum conditions, very consistent with the RAD measurements (ranging between ~200-325 μGy/day since landing) [20]. Consequently, we argue that even such a simple model could give us many useful answers

**3. Results**

*3.1 Solar Energetic Particle Events*

The LET by the September 1989 solar event versus depth in the water layer for no shielding material,

aluminum, polyethylene, water, regolith, carbon fiber and liquid hydrogen is shown in Fig. 2. All of the materials have the same area density of 10 g cm$^{-2}$.

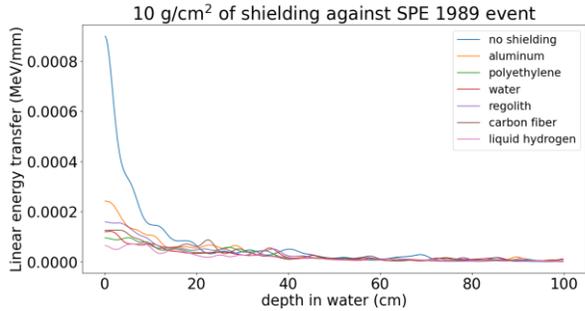

Fig. 2. LET by the September 1989 solar event versus depth in the water layer for no shielding material, aluminum, polyethylene, water, regolith, carbon fiber and liquid hydrogen. All of the materials have an area density of 10 g cm$^{-2}$. Figure reproduced from [17].

As expected, the lowest energy levels are provided by the liquid hydrogen shielding. However, we notice that the plots of the other materials are quite close. Among the ones we tested, aluminum appear to have the worst performance, which is attributed to the creation of a great number of secondary particles within its mass. Water, regolith, carbon fiber and polyethylene show an intermediate performance.

We continue by calculating the mean energy deposited inside the layer representing the astronaut. This is shown in Fig. 3. Results are given as a fraction compared to the dose with no shielding material. At a maximum radiation dose reduction of less than 60% for aluminum, one understands the importance of passive shielding materials.

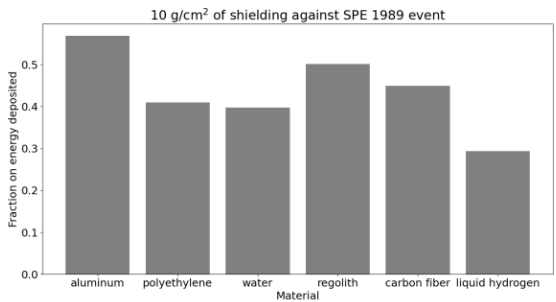

Fig. 3. Mean energy deposited in the water layer by the September 1989 solar event for aluminum, polyethylene, water, regolith, carbon fiber and liquid hydrogen. All of the materials have an area density of 10 g cm$^{-2}$. The values are given as a fraction compared to the no shielding situation.

*3.2 Galactic Cosmic Rays*

The respective results of LET and energy deposition, as generated by a GCR spectrum, are demonstrated in Fig. 4 and 5. We define a full GCR spectrum (protons, alpha particles, and iron ions as a representative for HZE ions) in minimum solar conditions (i.e., maximum GCR flux).

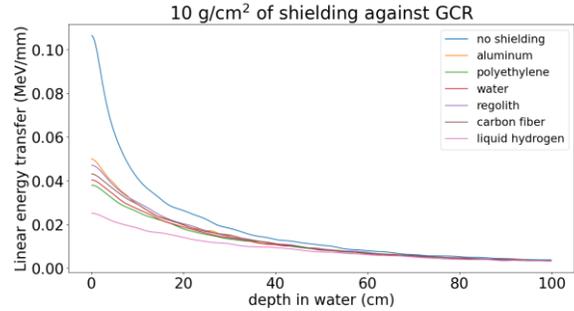

Fig. 4. LET by the GCR particles (minimum solar conditions) versus depth in the water layer for no shielding material, aluminum, polyethylene, water, regolith, carbon fiber and liquid hydrogen. All of the materials have an area density of 10 g cm$^{-2}$. Figure reproduced from [17].

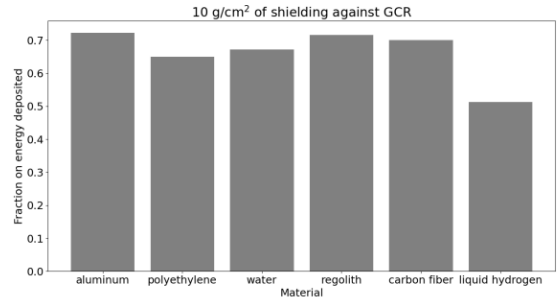

Fig. 5. Mean energy deposited in the water layer by the GCR particles (minimum solar conditions) for aluminum, polyethylene, water, regolith, carbon fiber and liquid hydrogen. All of the materials have an area density of 10 g cm$^{-2}$. The values are given as a fraction compared to the no shielding situation.

The results obtained by GCR particles are on general compatible with what we got from SEPs. Energies are of course larger though. Aluminum and regolith reduce the energy deposition by about 30%. Carbon fiber, water and polyethylene come next. At roughly 50% attenuation of the incoming radiation, liquid hydrogen ranks again as the most effective material. In spite of that, its poor mechanical properties make it difficult to be used in large scale.

Overall, polyethylene is considered the best shielding material, considering its practicability (mass efficiency). Its richness in hydrogen, along with its many electrons per unit mass, render it as a good shield against ions and their secondaries.

The properties of Martian regolith have a special interest. As it is an in-situ material, it does not require us to carry it from Earth, which is an advantage that can help a lot with the payload issues. Hence, an important question would be how regolith and combinations of it with other commonly used materials like aluminum, water or carbon fiber would respond against radiation. The results are shown in Fig. 6. We infer that, other than aluminum, the LET is lower for 10 g cm$^{-2}$ of a material compared to 5 g cm$^{-2}$ of regolith followed by 5 g cm$^{-2}$ of the material. However, differences are only small, and so, regolith is a good alternative. In order to maintain launch mass under reasonable limits, we recommend using Martian regolith as an extra shield against radiation while on Mars.

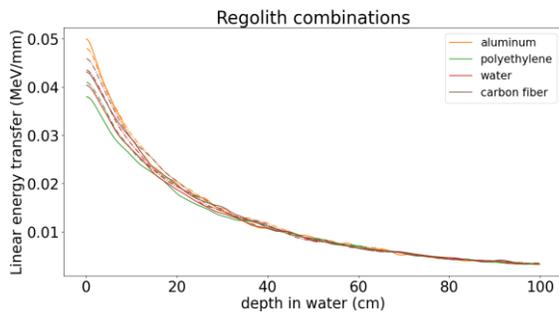

Fig. 6. LET by the GCR particles (minimum solar conditions) versus depth in the water layer for combinations of regolith (aluminum, polyethylene, water and carbon fiber). Solid lines represent 10 g cm$^{-2}$ of a material and dotted lines 5 g cm$^{-2}$ of regolith followed by 5 g cm$^{-2}$ of the material. Figure reproduced from [17].

Fig. 7 depicts the linear energy transfer induced by GCR ions (during a solar minimum period) through all simulated layers (1 to 6). We observe a major peak of energy deposition around 900 cm, which is inside the condensed atmospheric layer. This peak seems like a delta function, since the values of LET are a lot lower in the other layers.

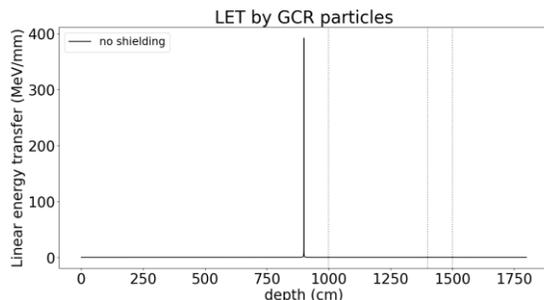

Fig. 7. LET by the GCR particles (minimum solar conditions) through all simulated layers. The vertical lines represent the limits of each layer.

## 5. Conclusions

To minimize radiation exposure, people living on Mars will need to adopt a series of strategies, the main being the passive shielding technique. Here, we examined some materials for their response again particles in the Martian environment. Our results indicate all tested materials reduced absorbed doses compared to not having any kind of shield. The most effective materials were the ones rich in hydrogen (polyethylene, water, or liquid hydrogen), but other materials like carbon fiber or aluminum still attenuate energy levels at a sufficient percent. Furthermore, we advocate using combinations of the in-situ Martian regolith with other materials, in such a way that we use each material's properties as an advantage.

### Acknowledgements

We thank the members of the Geant4 collaboration who developed the software (geant4.cern.ch) used for this work.